\documentclass[a4paper,fleqn,usenatbib]{mnras}

\usepackage[T1]{fontenc} \usepackage{ae,aecompl} \usepackage{graphicx}
\usepackage{graphics}     \usepackage{amsmath}    \usepackage{amssymb}
\usepackage{float}

\title[The   white  dwarf  population   of  the   Galactic  bulge]{The
  population of single and binary white dwarfs of the Galactic bulge}

\author[S.     Torres    et   al.]{S.     Torres$^{1,2}$\thanks{Email;
    santiago.torres@upc.edu},  E.   Garc\'ia--Berro$^{1,2}\dagger$, R.
  Cojocaru$^{1,2}$ and A.  Calamida$^{3}$\\ $^{1}$Departament de F\'\i
  sica, Universitat  Polit\`ecnica de Catalunya,  c/Esteve Terrades 5,
  08860  Castelldefels, Spain\\ $^{2}$Institute  for Space  Studies of
  Catalonia, c/Gran Capit\`a 2--4,  Edif.  Nexus 201, 08034 Barcelona,
  Spain\\  $^{3}$Space Telescope  Science Institute,  3700  San Martin
  Drive,  Baltimore,  MD   21218,  USA\\  $^{\dagger}$  Deceased  23rd
  September 2017}

\date{Accepted. Received; in original form}

\pubyear{2016}

\begin{document}
\label{firstpage}
\pagerange{\pageref{firstpage}--\pageref{lastpage}} \maketitle

\begin{abstract}
Recent  Hubble Space  Telescope observations  have unveiled  the white
dwarf cooling sequence of the Galactic bulge.  Although the degenerate
sequence can be well  fitted employing the most up-to-date theoretical
cooling  sequences,  observations  show  a systematic  excess  of  red
objects that cannot  be explained by the theoretical  models of single
carbon-oxygen white dwarfs of  the appropriate masses. Here we present
a population  synthesis study of  the white dwarf cooling  sequence of
the Galactic  bulge that  takes into account  the populations  of both
single white dwarfs  and binary systems containing at  least one white
dwarf.    These  calculations  incorporate   state-of-the-art  cooling
sequences for  white dwarfs with  hydrogen-rich and hydrogen-deficient
atmospheres,  for  both white  dwarfs  with  carbon-oxygen and  helium
cores,  and  also take  into  account  detailed  prescriptions of  the
evolutionary history of binary systems. Our Monte Carlo simulator also
incorporates all  the known observational  biases.  This allows  us to
model with a high degree of  realism the white dwarf population of the
Galactic bulge.  We find that the  observed excess of red stars can be
partially attributed  to white dwarf plus main  sequence binaries, and
to cataclysmic variables or dwarf novae. Our best fit is obtained with
a higher  binary fraction and  an initial mass function  slope steeper
than standard  values, as well  as with the inclusion  of differential
reddening  and  blending. Our  results  also  show  that the  possible
contribution  of double  degenerate  systems or  young and  thick-disk
bulge stars is negligible.
\end{abstract}

\begin{keywords}
stars: white dwarfs --  stars: binaries -- stars: luminosity function,
mass function -- Galaxy: bulge -- Galaxy: stellar content
\end{keywords}

\section{Introduction}

White  dwarfs  are the  most  abundant  fossil  stars in  the  stellar
graveyard. Actually,  it turns  out that more  than 90 percent  of all
stars  will end  their  evolution as  white  dwarfs.  Moreover,  their
structural and evolutionary  properties are reasonably well understood
-- see the  review of  \cite{Alt2010a} for an  in depth  discussion of
this  issue. Additionally,  supported  by the  pressure of  degenerate
electrons, white dwarfs slowly cool down for periods of time which are
comparable to the age of the Galaxy. All this allows us to explore the
evolution of various  stellar systems at early times.   As a matter of
fact, the properties of the white dwarf population have been used as a
valuable  tool  to study  the  nature  and  history of  the  different
components  of our  Galaxy --  including, for  instance, the  thin and
thick  disks \citep{Winget87, GB88a,  GB88b, GB1999,  MC2, Rowell2011,
  Rowell2013},   and   the   Galactic   halo   \citep{Mochkovitch1990,
  Isern1998,  MC3,  vanOirschot2014}  --  and the  characteristics  of
several  open  and  globular  clusters   --  of  which,  to  put  some
representative  and   recent  examples,   we  mention  the   works  of
\cite{Calamida2008},     \cite{GB2010},    \cite{2011ApJ...730...35J},
\cite{2013A&A...549A.102B},       \cite{2013Natur.500...51H}       and
\cite{2015A&A...581A..90T}.   Furthermore,  the  population  of  white
dwarfs carries  fundamental information about  several crucial issues,
like for  instance (but not only)  the amount of mass  lost during the
several evolutionary phases of  their progenitors -- see, for instance
the  recent review  of \cite{WDLF}.   In addition,  the  population of
white dwarfs  can be  used for other  important purposes.   We mention
here three of them.  First, the ensemble properties of the white dwarf
population  can be  used  to probe  the  behavior of  matter at  large
densities  and  low  temperatures \citep{1991A&A...241L..29I}.   White
dwarfs  can also  be used  to  understand the  evolution of  planetary
systems    across    the   evolutionary    life    of   their    stars
\citep{2016NewAR..71....9F}.   Finally, the statistical  properties of
the  population  of  Galactic  white  dwarfs  have  been  successfully
employed to corroborate or discard some non-standard physical theories
\citep{1992ApJ...392L..23I,  1995MNRAS.277..801G, 2008ApJ...682L.109I,
  2011JCAP...05..021G}.

\begin{figure}
\includegraphics[width=1.13\columnwidth]{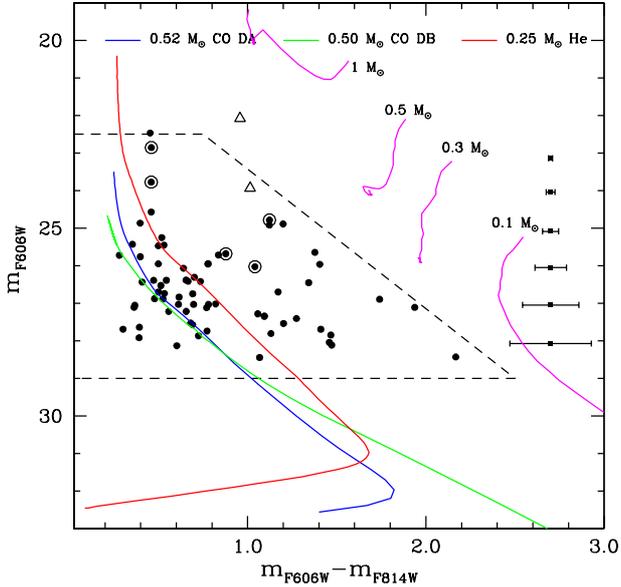}
   \caption{Color-magnitude     diagram    of     the     sample    of
     \protect\cite{Calamida2014}  --  black  symbols.   The  encircled
     points  are   cataclysmic  variable  candidates,   and  the  open
     triangles  are  confirmed nova-like  objects.   The dashed  lines
     indicate the selection region  of white dwarfs.  Also plotted are
     different  evolutionary  sequences   for  white  dwarf  and  main
     sequence stars as labeled. See text for details.}
\label{f:fig1}
\end{figure}

Modern large-scale  automated surveys, besides unveiling  many new and
interesting    astronomical   objects,    are    revolutionizing   our
understanding of the different Galactic populations. Specifically, the
Sloan Digital  Sky Survey \citep{2000AJ....120.1579Y},  the Pan-STARRS
collaboration    \citep{2002SPIE.4836..154K},    the    RAVE    Survey
\citep{2008AJ....136..421Z},    or   the   SuperCosmos    Sky   Survey
\citep{1998MNRAS.298..897H}, to cite a  few examples, have provided us
with  an unprecedented  wealth  of astrometric  and photometric  data.
Indeed,  both the  Sloan Digital  Sky Survey  and the  SuperCosmos Sky
Survey have  increased the number  of known white dwarfs  belonging to
the thin, thick and halo  populations.  Furthermore, a subset of these
enhanced  white dwarf  samples  has reliable  determinations of  their
observable  characteristics.  All  this  has allowed  us to  undertake
thorough  studies  of  these  three  Galactic  populations,  aimed  at
uncovering their main properties.

However, because of the lack  of observational data, this has not been
possible  yet for  the Galactic  bulge.  Recently,  a first  sample of
white  dwarfs  belonging  to  the  Galactic bulge  has  been  released
\citep{Calamida2014}.    Observations  of   the   {\sl  Hubble   Space
  Telescope}  towards  the  low-reddening  Sagittarius window  in  the
Galactic  bulge were deep  enough to  measure accurate  magnitudes and
proper motions for a sample  of white dwarfs, allowing us to determine
a   relatively   clean    color-magnitude   diagram   of   the   bulge
\citep{Calamida2014}.   The  white  dwarf  cooling  sequence  contains
$\sim70$ white dwarfs  candidates.  Moreover, \cite{Calamida2014} also
identified  $\sim20$ objects  among extreme  horizontal  branch stars,
dwarf  novae and cataclysmic  variables.  However,  a large  spread in
color  is  observed  with  $\sim30\,\%$  of the  objects  redder  than
expected for a typical carbon-oxygen white dwarf.  This indicates that
presumably there  may be a  substantial fraction of white  dwarfs with
helium cores, whose  nature and origin remain to  be elucidated.  Here
we analyze, with  the help of a Monte  Carlo population synthesis code
the observational  data of this  sample of stars, taking  into account
all the known observational biases and restrictions. We explore, among
other issues,  the nature of  the observed population of  red objects,
the fraction of  binaries and the population of  double degenerates in
the Galactic bulge.

Our  paper is  organized as  follows.  In  Sect.~\ref{observations} we
describe  with  some detail  the  set  of  observations to  which  our
theoretical  simulations will be  compared.  Our  numerical set  up is
briefly described in Sect.~\ref{setup}. The results of our Monte Carlo
population  synthesis calculations  are  described in  full detail  in
Sect.~\ref{results}. In  particular, in Sect.~\ref{single}  we discuss
the  properties  of  the   population  of  single  white  dwarfs  with
carbon-oxygen  cores.   Sect.~\ref{helium} is  devoted  to assess  the
possible existence of a  sub-population of single helium white dwarfs,
while in Sect.~\ref{bin} we analyze  the contribution of the sample of
binary systems made of a carbon-oxygen white dwarf and a main-sequence
star.  Other  minor effects,  differential reddening and  blending are
studied in Sect.~\ref{other}. Finally, Sect.~\ref{sec:conc} is devoted
to summarize our main results and to discuss their significance.

\section{The observed color magnitude diagram}
\label{observations}

In  Fig.~\ref{f:fig1} we  present the  color-magnitude diagram  of the
sample of \cite{Calamida2014}. The  black points denote objects of the
observed  population,  the  open  triangles  are  confirmed  nova-like
objects, and  the black encircled  points correspond to  those objects
that,    according   to   \cite{Calamida2014}    present   photometric
variability,  and presumably  are cataclysmic  variables.   The dashed
lines  delimit   the  selection  region   for  white  dwarfs   in  the
color-magnitude diagram.  In particular, the bottom line indicates the
limiting magnitude of the survey  where the completeness of the sample
drops  below  a  $50\%$.  The  upper  line  is  the magnitude  of  the
brightest   white  dwarfs   according  to   the  cooling   tracks  for
carbon-oxygen white dwarfs at the distance and reddening of the bulge,
while the  oblique line represents the main-sequence  blue edge.  Also
displayed in Fig.~\ref{f:fig1} are  the theoretical cooling tracks for
a  $0.52\,{M_{\sun}}$ hydrogen-rich white  dwarf with  a carbon-oxygen
core   \citep{Renedo2010}  --  blue   line  --   a  $0.50\,{M_{\sun}}$
carbon-oxygen  white   dwarf  with  a   hydrogen-deficient  atmosphere
\citep{Benvenuto1997} -- green line  -- and a $0.25\,{M_{\sun}}$ white
dwarf with helium core \citep{Serenelli2001} -- red line.  These white
dwarf  masses correspond  to  the turn-off  mass  of the  bulge for  a
typical age $\sim 11$~Gyr.  Clearly, although white dwarfs with helium
cores  are redder than  regular carbon-oxygen  white dwarfs,  a region
populated by even  redder objects exits. Thus, objects  in this region
cannot  be  explained  by  the most  up-to-date  evolutionary  stellar
tracks,  even  when  photometric  errors are  considered.   A  natural
explanation  for  these stars  is  that  they  may be  binary  systems
composed of a white dwarf and a main sequence star (WDMS).  To address
this issue,  and only for illustrative  purposes, in Fig.~\ref{f:fig1}
we have also plotted  several evolutionary sequences for low-mass main
sequence stars  (magenta lines) computed  by \cite{Baraffe2015}. These
evolutionary  sequences are  plotted to  show that  the presence  of a
low-mass  companion in  a non-resolved  binary system  made of  a main
sequence star and a white  dwarf results in redder colors and brighter
objects.   Finally,  note  as  well  that within  the  region  in  the
color-magnitude  diagram  defined by  the  selection  region of  white
dwarfs of the survey, the cooling tracks of white dwarfs with hydrogen
rich and hydrogen deficient atmospheres do not differ by much.

\begin{figure}
\includegraphics[width=1.13\columnwidth]{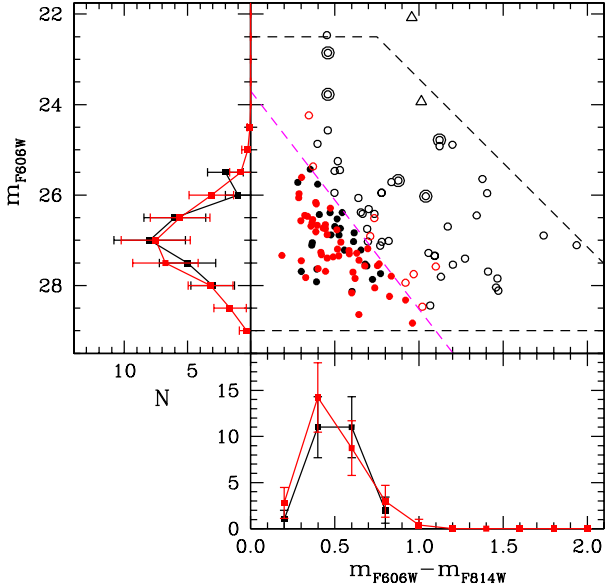}
   \caption{Color-magnitude diagram  (central panel) for  the observed
     sample of \protect\cite{Calamida2014} -- black symbols -- and for
     our simulated  sample for the  Galactic bulge single  white dwarf
     population -- red symbols. We assumed that objects bluer than the
     dividing line (magenta dashed line) are single white dwarfs.  The
     corresponding   luminosity  function   (left  panel)   and  color
     distribution (bottom  panel) for the  observational (black lines)
     and simulated sample (red lines) are also shown. Excluded objects
     are  marked  with  open   symbols,  while  encircled  points  are
     cataclysmic variable candidates  and open triangles are confirmed
     nova-like objects. }
\label{f:fig2}
\end{figure}

\section{Building the synthetic sample}
\label{setup}

The Monte  Carlo population synthesis simulator employed  in this work
has  been   successfully  used  in  several   occasions.   A  thorough
description  of  its  main  inputs  can  be  found  in  \cite{GB1999},
\cite{MC2} and  \cite{MC3}.  Consequently, here we  will only describe
its most relevant inputs, and  we refer the interested reader to these
works for detailed information of the rest of ingredients of the Monte
Carlo code.  For the calculations reported below, and except otherwise
explicitly  stated, we  define a  reference model  with  the following
inputs. The  mass of synthetic  main sequence stars is  drawn randomly
according  to a Salpeter-like  initial mass  function with  a standard
slope,  $\alpha=-2.35$,  but  below  we explore  other  choices.   The
adopted  age  for   our  fiducial  model  of  the   bulge  is  $T_{\rm
  c}=11.0$~Gyr  with a constant  burst of  star formation  lasting for
$1$~Gyr.   An alternative  model  with a  younger  population is  also
studied  in Section \ref{young}.   The population  of bulge  stars has
three  sub-components of  metallicities $Z=0.03$,  $0.02$  and $0.008$
\citep{Haywood2016}.  We use a  set of evolutionary sequences covering
the full range of masses  and metallicities of the progenitor stars of
the   white   dwarf   population   \citep{Renedo2010,   Benvenuto1997,
  Serenelli2001}.  The set of  cooling sequences for white dwarfs with
pure hydrogen atmospheres was evolved self-consistently from the ZAMS,
through  the giant  phase,  the thermally  pulsing  AGB and  mass-loss
phases, and  ultimately to the  white dwarf cooling phase.   All these
sets of cooling sequences cover  the full range of masses of interest.
In  the case  of  white  dwarf-main sequence  binaries,  for the  main
sequence   companion   we   employ   the  evolutionary   tracks   from
\cite{Baraffe2015}.  These evolutionary sequences are valid for masses
ranging from  0.07 to $0.5\,M_{\sun}$.  A fraction,  $f_{\rm BIN}$, of
the total  mass generated is evolved through  binary systems following
the   prescriptions  of   \cite{Hurley2002},  with   the  improvements
described in  \cite{Camacho2014} and \cite{Cojocaru2017}.   Colors and
magnitudes  are transformed  from  the Johnson-Cousins  system to  the
VEGAMAG   zero  point   magnitude  system   using  the   procedure  of
\cite{2005PASP..117.1049S}.   In our simulations  we adopt  a distance
modulus       $(m-M)_{\rm       0}=14.45\,$mag,      a       reddening
$E(F606W-F814W)=0.5766$ and  an absorption $A_{\rm F606W}=1.4291\,$mag
\citep{Calamida2014}.  The fraction of white dwarfs with hydrogen-rich
atmospheres is  80\%, the  canonical choice, while  20\% of  stars are
adopted  to  be  white  dwarfs with  hydrogen  deficient  atmospheres.
Finally,  in order to  strictly mimic  the observational  procedure we
implemented the corresponding magnitude  and color cuts of the survey,
added  the  reported photometric  errors  and  took  into account  the
completeness of the sample.

\section{Results}
\label{results}

\subsection{The population of single white dwarfs}
\label{single}

To start with, we discuss the population of single white dwarfs of the
Galactic bulge.  For this analysis  we selected stars in the sample of
\cite{Calamida2014}   that  have   photometry   compatible  with   the
expectations for  single white dwarfs with  carbon-oxygen cores.  This
was  done by  considering the  region of  the  color-magnitude diagram
where  a  typical $0.52\,{M_{\sun}}$  white  dwarf with  hydrogen-rich
atmosphere and  with a carbon-oxygen  core would be  eventually found.
This is  the mass of  the white dwarf  for a progenitor with  the main
sequence turn-off mass of $\approx 0.95\,M_{\sun}$.  The corresponding
cooling track for  this mass is shown in  Fig.~\ref{f:fig1} as a blue,
solid line.  We then  added a conservative $1\sigma$ photometric error
to this cooling  sequence to obtain the region  in the color-magnitude
diagram  where these  white dwarfs  should be  located.   Objects with
redder colors  should be binary  systems, or white dwarfs  with helium
cores, whatever  their evolutionary origin  could be.  After  this, we
computed  the  magnitude  distribution   (that  is,  the  white  dwarf
luminosity function) and the  color distribution.  The results of this
procedure are shown in  Fig.~\ref{f:fig2}.  In particular, the central
panel of this figure displays  the color magnitude diagram of both the
observed sample  and of the  simulated sample.  Again, in  this figure
the dashed black  lines indicate the selection region  of white dwarfs
of the  survey.  The  dashed magenta line  divides the regions  in the
color-magnitude diagram  where single  white dwarfs should  be located
(to  the left of  this line)  and where  with the  highest probability
binary systems or  white dwarfs with helium cores  are expected to lay
(to the right  of this line).  Thus, filled  red symbols are synthetic
white dwarfs with carbon-oxygen cores and filled black symbols are the
observed sample  of white dwarfs with  presumably carbon-oxygen cores.
Open  black symbols correspond  to observed  white dwarfs  most likely
having helium or  belonging to binary systems, and  hence are not used
to  obtain   the  white  dwarf  luminosity  function   and  the  color
distribution -- see below.   Finally, encircled symbols indicate those
objects that were reported to be variable by \cite{Calamida2014}.  Now
we turn  our attention to the  left panel of this  figure, which shows
the observed  (black line)  and the synthetic  (red line)  white dwarf
luminosity functions.   As can be  seen, despite the scarce  number of
observed stars,  the agreement between the simulated  and the observed
sample  is  excellent.  The  same  can  be  said regarding  the  color
distribution,  see the  bottom panel  of Fig.~\ref{f:fig2}.   Thus, we
conclude that our reference model, which we remind was computed with a
canonical ratio of 80\% of  white dwarfs with hydrogen atmospheres and
20\% of white dwarfs with hydrogen-poor atmospheres, reproduces fairly
well the population of single white dwarfs.

\begin{figure}
\includegraphics[width=1.13\columnwidth]{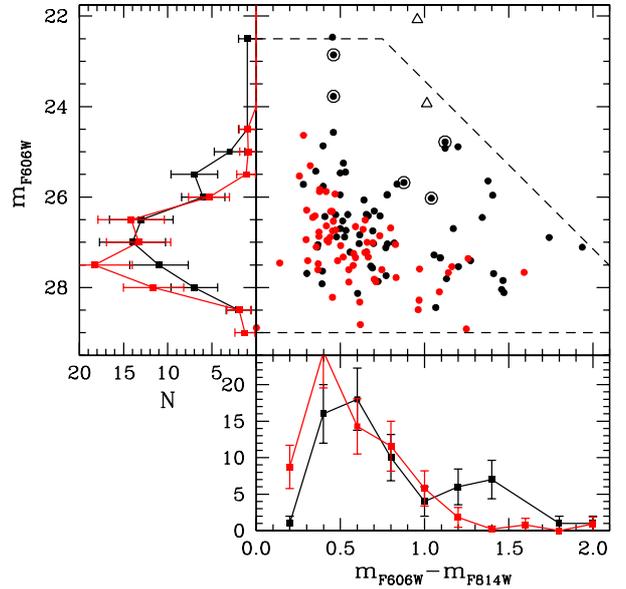}
   \caption{Same as Fig.~\ref{f:fig2} but for the population of single
     and  binary white dwarfs  of the  Galactic bulge.   Our simulated
     sample (red symbols and red lines) corresponds to Model~1.}
\label{f:mod1}
\end{figure}

\begin{table*}
\caption{Models  of the binary  population used  in this  work.  Lasts
  columns show  the final percentages obtained for  the populations of
  single   white   dwarfs,   double   degenerates  and   WDMS   pairs,
  respectively,   and  the  reduced   $\chi_{\nu}^2$  value   for  the
  luminosity function and the color distribution, respectively.}
\begin{center}
\begin{tabular}{ccccccccccc}  
\hline \hline Model & $f_{\rm BIN}$ & $\alpha_{\rm CE}$ & $\alpha_{\rm
  int}$ &  $n(q)$ & $\alpha_{\rm IMF}$  & \% WD &  \% DD &  \% WD+MS &
$\chi^2_{\nu,{\rm LF}}$ & $\chi^2_{\nu,{\rm CD}}$  \\ \hline 0 & 0.5 &
0.3 & 0.0 & 1 & $-2.35$ & 97.5 &  0.0 & 2.5 & 1.13 & 2.13 \\ 1 & 0.8 &
0.3 & 0.0 & 1 & $-2.35$ & 90.3 &  1.4 & 8.3 & 1.05 & 2.14 \\ 2 & 0.8 &
0.3 & 0.0 & 1 & $-2.55$ & 90.0 &  0.5 & 9.5 & 1.09 & 1.91 \\ 3 & 0.8 &
0.3 & 0.0 & 1 & $-2.15$ & 89.0 &  2.8 & 8.2 & 1.15 & 2.13 \\ 4 & 0.8 &
1.0 & 0.1 & 1 & $-2.35$ & 90.1 &  1.4 & 8.4 & 1.10 & 2.06 \\ 5 & 0.8 &
0.3 & 0.0 & $q$ & $-2.35$ & 94.1 &  0.0 & 5.9 & 1.11 & 1.88 \\ 6 & 0.8
& 0.3  & 0.0 & $q^{-1}$ &  $-2.35$ & 82.5 &  2.0 & 15.5 &  0.87 & 1.81
\\ 7 & 0.8 & 1.0 & 0.1 &  Duquennoy & Kroupa & 87.8 & 2.7 & 9.5 & 0.83
& 1.61 \\ & & & \& Mayor\,(1991) & et  al. (2001) & & & & & \\ 8 & 0.8
& 1.0  & 0.1 & $q^{-1}$ &  $-2.55$ & 77.9 &  1.0 & 21.1 &  1.09 & 1.17
\\ \hline \hline
\end{tabular}
\end{center}
\label{t:table1}
\end{table*}

\subsection{The population of white dwarfs with helium cores}
\label{helium}

\begin{figure}
\includegraphics[width=1.13\columnwidth]{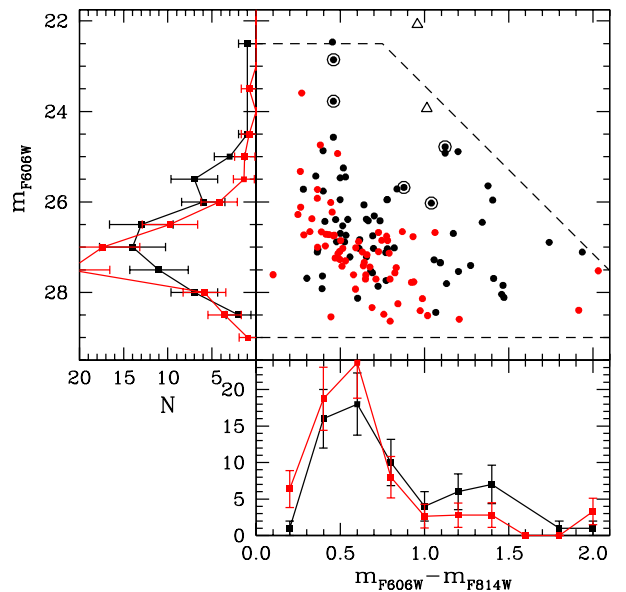}
   \caption{Same as Fig.~\ref{f:fig2} but for the population of single
     and  binary white dwarfs  of the  Galactic bulge.   The simulated
     sample (red symbols and red lines) is that of Model~8.}
\label{f:mod8}
\end{figure}

Now we turn  our attention to assess the existence  of a population of
single white  dwarfs with helium  cores. There is a  general agreement
that the  vast majority  of low-mass white  dwarfs, namely  those with
$M_{\rm    WD}    \leq     0.47\,{M_{\sun}}$,    have    evolved    in
binaries. Otherwise their low-mass main sequence precursors would need
more  than the  Hubble time  to  leave the  main sequence.   Moreover,
\cite{RM2011} analyzed  a large sample  of short orbital  period WD+MS
binaries  that evolved through  a common  envelope phase  and provided
robust  observational  evidence that  the  majority  ($\sim 85\%$)  of
low-mass white dwarfs were formed as a consequence of mass transfer in
binaries.   However,  the  origin  of  the remaining  $\sim  15\%$  of
low-mass white  dwarfs still lacks a  consistent explanation.  Several
scenarios  have  been  proposed.   In particular,  Type  Ia  supernova
explosions in semi-detached close binaries \citep{Justham2009}, severe
mass-loss     during     the      first     giant     branch     phase
\citep{Kilic2007,Meng2008},  ejection of the  stellar envelope  due to
the  spiral-in of  close  giant planets  \citep{Nelemans1998}, or  the
merging of two extremely  low-mass white dwarfs \citep{Han2002}, among
other possible scenarios, have  been proposed.  Whatever the origin of
these single white dwarfs with  helium cores could be, they are redder
than  typical white  dwarfs  with carbon-oxygen  cores.  This  becomes
clear by examining  Fig.\ref{f:fig1}, where we show using  a solid red
line  the cooling  track  for a  $0.25\,{M_{\sun}}$ helium-core  white
dwarf.

To  evaluate the  impact  of these  white  dwarfs in  the white  dwarf
luminosity function and  on the color distribution of  the white dwarf
population of the  Galactic bulge we conducted a  suite of simulations
where we included  a variable fraction, $f_{\rm He}$,  of white dwarfs
with helium cores.  For an age  of the bulge of $11\,$Gyr and adopting
Solar metallicity, the turn-off mass is $\approx1\,{M_{\sun}}$ and the
maximum mass of the progenitors that could have produced a helium-core
white  dwarf  is  $\approx2\,{M_{\sun}}$.   Thus, we  assumed  that  a
fraction $f_{\rm  He}$ ranging from $20\%$ to  $50\%$ of main-sequence
stars with masses in the interval between $1$ and $2\,{M_{\sun}}$ when
they reach  the giant  branch produce a  helium-core white  dwarf with
masses within  the interval $0.25$ and  $0.45\,{M_{\sun}}$.  Even when
the  most extreme  assumption  is  adopted, that  is  when a  fraction
$f_{\rm He}=0.50$  is employed, our Monte Carlo  simulations show that
the  probability of  finding a  single helium-white  dwarf  within the
observed sample is far  below $10^{-3}$.  Consequently, these stars do
not  contribute significantly  to the  number counts,  and  thus their
impact   on  the   magnitude  and   color  distributions   is  totally
negligible. In principle, we  need to consider that several non-linear
causes  are  present  in  this   final  low  probability,  such  as  a
non-constant initial mass function or a spread in born-ages due to the
1 Gyr  burst.  However,  we point  out that the  main reason  for this
result is  that helium-core  white dwarfs cool  down much  faster than
regular  white dwarfs  with carbon-oxygen  cores.  In  particular, the
time needed  for a typical $0.25\,{M_{\sun}}$  helium-core white dwarf
to cross the region within the selection limits of white dwarfs of the
survey -- which correspond  to effective temperatures between $\approx
27,500\,$K  and $\approx 4,500\,$K  -- is  less than  $\sim 4.4\,$Gyr,
while for a typical $0.6\,{M_{\sun}}$ white dwarf with a carbon-oxygen
core is $\sim 9.1\,$Gyr.

\subsection{The binary white dwarf population}
\label{bin}

The analysis  done in  the previous section  also demonstrates  that a
significant fraction of objects in the observed sample -- namely those
with colors  redder than usual  -- are probably binary  systems.  This
percentage is $\sim63\%$, a high  value.  To verify if this hypothesis
is correct, we assumed that a fraction $f_{\rm BIN}$ of the total mass
generated in the theoretical simulations goes to form binaries, and we
kept  all other  inputs fixed,  including the  initial  mass function.
Starting with a canonical value of $f_{\rm BIN}=0.50$ -- we refer this
model  as Model~0  -- we  increased $f_{\rm  BIN}$ so  that  the final
percentage  of binaries is  maximize but  keeping a  good fit  for the
luminosity function  and color distribution of the  single white dwarf
sample.   Specifically,  we  assumed  that this  fraction  is  $f_{\rm
  BIN}=0.80$. Additionally  we explored  other choices of  the initial
mass  function, as  well  as other  options  for the  age spread.   We
assumed that  the initial distribution  of mass ratios of  the primary
and secondary stars, $n(q)$ where  $q=M_2/M_2$ and $M_1$ and $M_2$ are
the  masses  of  the  primary   and  secondary  members  of  the  pair
respectively,  is flat.   We also  adopted  a value  for the  envelope
efficiency $\alpha_{\rm CE}=0.3$, with no internal energy contribution
$\alpha_{\rm   int}=0.0$,   in   accordance   with  the   results   of
\cite{Camacho2014}, and with all these inputs we generated a sample of
synthetic  binaries.  From  now  on we  will  refer to  this model  as
Model~1.

In  Fig.~\ref{f:mod1} we  present the  results of  this model,  and we
compare them  with the full  sample of \cite{Calamida2015}. As  we did
previously  we employ  black  symbols to  represent the  observational
data. We  emphasize that now  the white dwarf luminosity  function and
the color distribution -- see  below -- have been computed taking into
account all the white dwarfs observed by \cite{Calamida2014}, not only
those white dwarfs that are single and with carbon-oxygen cores, as we
did previously  (in Sect.~\ref{single}).  Comparing Figs.~\ref{f:fig2}
and  \ref{f:mod1} it is  quite apparent  that the  observed luminosity
function (left  panel, black line) and the  color distribution (bottom
panel, black line)  of the full sample of  white dwarfs have secondary
peaks  that were  not present  in  the luminosity  function and  color
distribution of Fig.~\ref{f:fig2}.   These secondary peaks are located
at magnitude  $m_{\rm F606W}\approx 25.5$ for  the luminosity function
and at color $(m_{\rm  F606W}-m_{\rm F814W})\approx 1.4$ for the color
distribution, respectively.  Naturally, this is due to the red objects
that  we did  not consider  in  our previous  analysis.  Finally,  the
synthetic stars resulting from Model~1  are shown in the central panel
of  Fig.~\ref{f:mod1} using  red  symbols.  Examining  this figure  is
evident that Model~1, that we remind considers the population of WDMS,
is  not  able   to  reproduce  the  main  features   of  the  observed
color-magnitude  diagram. Consequently,  the  theoretical white  dwarf
luminosity function (left panel,  red line) and the color distribution
(bottom  panel, red  line) do  not match  the  observed distributions.
Specifically,   Fig.~\ref{f:mod1}   clearly   demonstrates  that   the
synthetic population of Model~1 does not reproduce the secondary peaks
of the  observed distributions.  Indeed, although  this model produces
some objects that are redder than usual, the fraction of these objects
is $8.3\%$, far below the  percentage of these objects in the observed
sample.

Thus, an alternative  model must be sought. Consequently  we explore a
suite of models aimed to obtain  a larger fraction of WDMS pairs.  The
different models studied here are listed in Table~\ref{t:table1}.  The
first column  displays the model,  being Model~1 our  reference model.
In  the next  columns the  values of  $\alpha_{\rm  CE}$, $\alpha_{\rm
  int}$, $n(q)$, $\alpha_{\rm IMF}$, and the percentages of the single
stars,  double degenerate  binaries and  WDMS pairs  of  the synthetic
sample are  listed. For a  quantitative comparison of the  models, the
reduced $\chi_{\nu}^2$  test value of the luminosity  function and the
color      distribution      was      calculated.      Defined      as
$\chi_{\nu}^2=\chi^2/\nu$ where $\nu$ is  the degree of freedom of the
respective function,  the corresponding results are shown  in the last
two  columns  of  Table~\ref{t:table1}.   Models  2 to  6  are  slight
variants of Model~1. For these  models only one parameter is modified.
In particular, Models~2 and 3 are intended to explore the slope of the
initial mass  function.  In  Model~4 we explore  the role of  the free
parameters of the common envelope  phase.  Finally, Models~5 and 6 are
thought to  explore the  effects of $n(q)$,  the distribution  of mass
ratios within the binary system.   Model~7 is the model that best fits
the  mass function  of  the Galactic  bulge \citep{Calamida2015}.   In
particular, we use for this  model the log-normal distribution of mass
ratios of  binary stars of  \cite{Duquennoy1991} and the  initial mass
function  of \cite{Kroupa2001}. In  short, the  results shown  in this
table show that  larger percentages of WDMS pairs  are produced when a
decreasing binary  mass ratio  distribution is adopted  (Model~6), and
when  a larger  common envelope  efficiency with  internal  energy are
adopted.   Also, we find  that a  steeper slope  for the  initial mass
function    fits    better    the    distribution   of    masses    of
\cite{Calamida2015}.   Finally, Model~8  is tailored  to  maximize the
percentage of synthetic binary systems.  For this particular model the
percentage of  binary systems is  the largest, $21\%$. Besides  we can
check  that the  reduced $\chi^2$  value is  the closest  to 1  of all
models,  thus concluding  that  Model~8 fits  better  the white  dwarf
luminosity  function and  the color  distribution.   The corresponding
luminosity  function, magnitude-diagram  color and  color distribution
are shown in Fig.~\ref{f:mod8}.   However, our simulations also reveal
that  there is  a  region  of the  color-magnitude  diagram where  our
simulations  do not  yield any  binary system,  at odds  with  what is
observationally found.

\begin{figure}
\includegraphics[trim       =      40mm       5mm       20mm      5mm,
  clip,width=1.4\columnwidth]{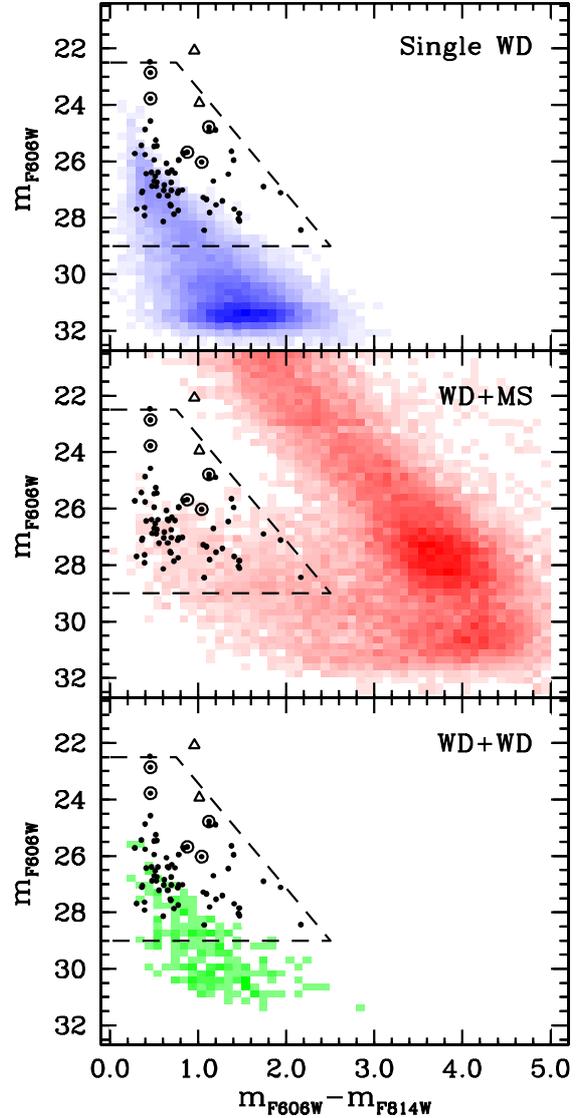}
   \caption{Density map  of the  Galactic bulge populations  of single
     white dwarfs (top panel), white dwarf plus main-sequence binaries
     (central panel) and double-degenerate systems (bottom panel). The
     results plotted  here are those  of our Model~8,  which maximizes
     the fraction of binaries.}
\label{f:map3}
\end{figure}

To study  which systems may eventually  be found in the  region of the
color-magnitude diagram that our simulations are not able to populate,
in  Fig.~\ref{f:map3}  we show  density  maps  of  the region  of  the
magnitude-color diagram  populated by  the population of  single white
dwarfs  (top  panel), the  WDMS  population  (central  panel) and  the
population of double degenerate systems (bottom panel), obtained using
Model~8.  It  is interesting  to note that  the bulk of  the different
subpopulations lays outside of  the observed region. For instance, the
vast majority  of single  white dwarfs have  cooled down  to magnitude
$m_{\rm F606W}\approx  31.5$, far below the completeness  limit of the
sample. On  the other  hand, WDMS systems  have magnitudes  within the
selection region of  white dwarfs of the survey,  however their colors
are   considerably   redder   than   the  observational   color   cut.
Consequently,   only  objects   of   the  respective   tails  of   the
distributions  are  culled.    Finally,  the  contribution  of  double
degenerate  systems is quite  marginal and  can be  safely disregarded
given that they  represent no more than a $2\%$ of  the systems in the
observable window.

\begin{figure}
\includegraphics[width=1.13\columnwidth]{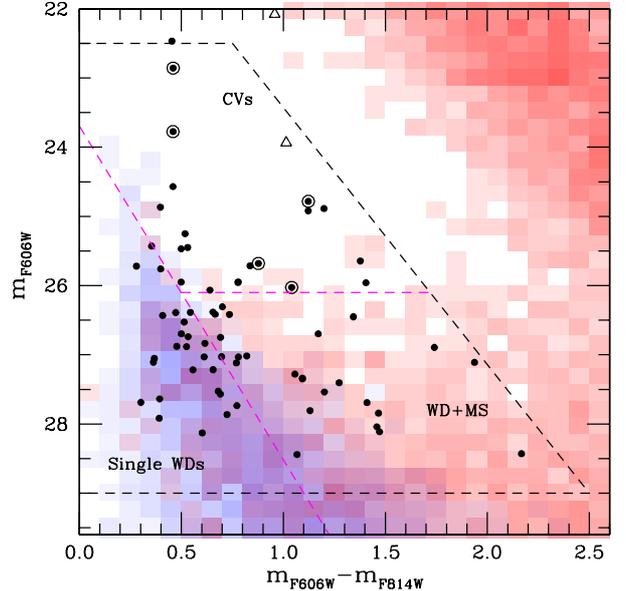}
   \caption{Same as Fig.~\ref{f:fig2}, but combining the population of
     single and binary white dwarfs of the Galactic bulge. The magenta
     lines indicate the regions  where the populations of single white
     dwarfs   and  WDMS  systems,   respectively,  are   dominant,  as
     labelled. The results presented in this figure are those obtained
     using Model~8.}
\label{f:map1}
\end{figure}

In Fig.~\ref{f:map1}  a density  map of the  observable region  of the
color-magnitude diagram for  the entire synthetic population, weighted
by their relative contribution to the final sample is shown. As can be
seen, within the observable  window of the color-magnitude diagram the
different  populations  overlap.   Clearly,  the  observed  sample  is
dominated by the single white dwarf sub-sample (blue color) and by the
WDMS population  (red color), while  the contribution of  double white
dwarf systems is negligible.  Taking into account which is the largest
percentage of  the different populations the observable  region can be
divided   in   smaller  regions.    The   magenta   dashed  lines   in
Fig.~\ref{f:map1} show  the limits  of these regions.   Basically, the
region  explored  by  \cite{Calamida2014}  can  be  divided  in  three
regions.  Objects with colors $m_{\rm F606W}-m_{\rm F814W}\la 1.0$ and
magnitudes   ranging  from   $m_{\rm  F606W}\simeq   29$  to   24  are
predominantly single  white dwarfs.  Observed  objects with magnitudes
between 29 and  $\sim 26$ and colors redder than  those of the magenta
dashed line  are most likely WDMS  systems.  The only  region that our
simulations are not able to  cover are objects brighter than magnitude
26 and colors redder than that  of the magenta dashed line. This means
these objects should be cataclysmic variables or dwarf novae, since we
are left with no other  option.  Moreover, it is worth mentioning that
\cite{Calamida2014}  found convincing  evidence for  variability  of a
sizable fraction  of objects in  this region.  Although we  cannot use
this approach to assign a  membership to each individual object in the
sample of  \cite{Calamida2014}, we can roughly estimate  the nature of
the observed sample.  According  to our simulations, $\sim40\%$ of the
observable  sample are  probably single  white dwarfs,  $\sim38\%$ are
WDMS systems and nearly  $\sim22\%$ are cataclysmic variables or dwarf
novae.

\begin{figure}
\includegraphics[trim      =       40mm      35mm      20mm      25mm,
  clip,width=1.35\columnwidth]{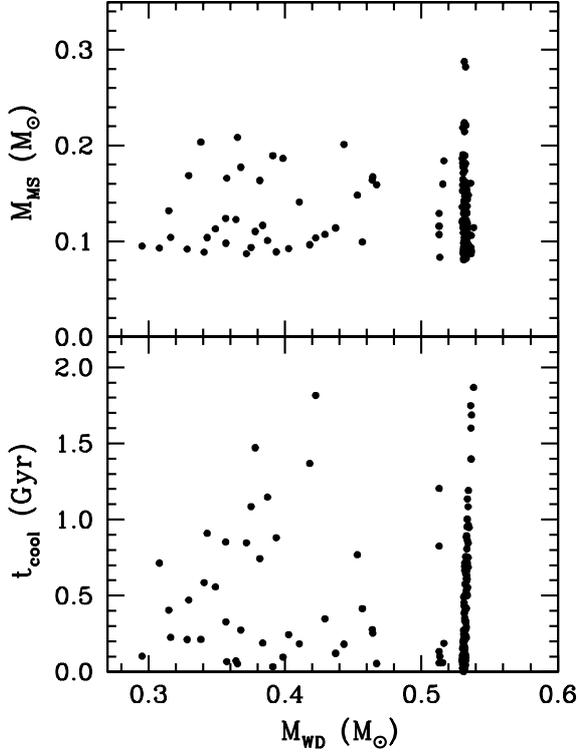}
   \caption{Main-sequence mass  as a function of the  white dwarf mass
     (top panel) and  cooling age of the white dwarf  as a function of
     its mass  (bottom panel) for those  WDMS systems that  lay in the
     observational  window of  \protect\cite{Calamida2015}.   The data
     presented in this figure are those resulting from Model~8.}
\label{f:wdms}
\end{figure}

Finally, we discuss the properties  of the population of WDMS binaries
that  contribute  to  the  observed  sample.   In  the  top  panel  of
Fig.~\ref{f:wdms}  we show the  mass of  the main  sequence star  as a
function of the mass of the white dwarf companion.  As can be seen two
formation channels coexist. The main channel, which accounts for $\sim
67\%$  of  the   binary  systems,  produces  a  $\sim0.54\,{M_{\sun}}$
carbon-oxygen  white dwarf and  a low-mass  main-sequence star  with a
mass ranging from $0.1$  to $0.3\,{\rm M_{\sun}}$. The second channel,
which accounts for the rest of  the binary systems, results in a white
dwarf   with   a   helium    core   and   mass   between   $0.3$   and
$0.44\,{M_{\sun}}$, and  a low-mass main-sequence  companion with mass
in  the interval  $0.1$  to $0.2\,{M_{\sun}}$.   Naturally, this  also
results in different cooling  times for both formation channels.  This
is illustrated in the bottom panel of the same figure, where the white
dwarf cooling times of the member  of the pair are shown as a function
of its mass.  It is worth emphasizing that for both formation channels
the white dwarf  cooling ages are shorter than  2~Gyr.  This indicates
that the WDMS  systems must be made of relatively  hot white dwarfs to
enter into the observational sample, otherwise the color of the system
would be dominated by the main-sequence companion, and consequently it
would be redder.

\subsection{A young bulge population?}
\label{young}

The  origin of  the  bulge still  remains  controversial. While  there
exists  a  general consensus  that  the  vast  majority of  the  bulge
population is uniformly old, there  is some evidence for a significant
younger population --  see \cite{Haywood2016}, and references therein.
The  observed  bar-like  shape  of  the Galaxy  seems  to  be  clearly
connected to the thick  disk structure and, consequently, the observed
dispersion in  metallicities of the  bulge stars can be  associated to
the different stellar  populations of the Galaxy \citep{DiMatteo2016}.
To check whether a younger  bulge population has noticeable effects on
the results  presented until  now we ran  an additional  simulation to
take into  account a younger  population.  In particular,  we slightly
modified  Model~8  to include  a  star  formation  rate history  which
incorporates  a young population.   We adopt  a simplified  model that
reasonably resembles  that of  \cite{Haywood2016} -- see  their Fig.~1
for  details.  That is,  to the  burst of  star formation  of duration
1~Gyr  that  occurred  11.0~Gyr   ago  and  produced  stars  with  low
metalicity  we added a  constant formation  period, lasting  until the
present, of  metal-rich stars  with a formation  rate 20\% of  that of
that of the burst.

\begin{figure}
\includegraphics[width=1.13\columnwidth]{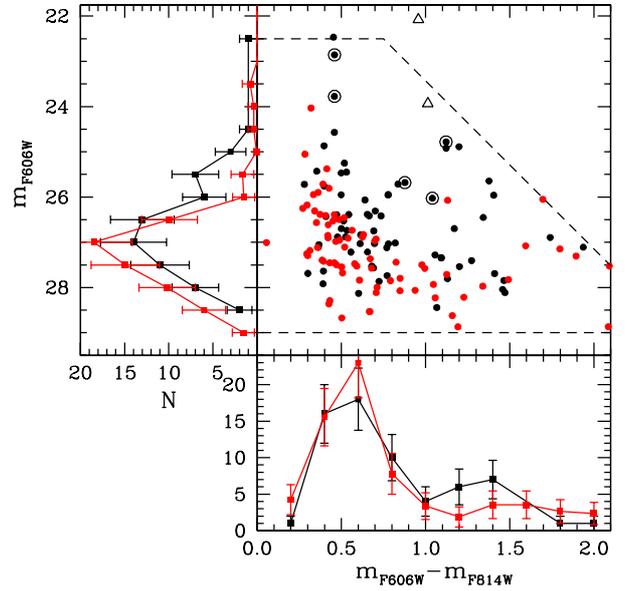}
   \caption{Same as Fig.~\ref{f:mod8} but for a star formation history
     that includes a young bulge population.}
\label{f:mod8sfr}
\end{figure}

The results of employing this  star formation history are displayed in
Fig.~\ref{f:mod8sfr}.  As  in Fig.~\ref{f:mod8} we  show our simulated
sample  (black  symbols  and   black  solid  lines)  compared  to  the
observational dataset  (red dots  and solid red  lines). First,  it is
important  to mention  that including  a younger  population  of stars
substantially  increases the  number  of WDMS  binaries  in the  final
sample. Specifically we  obtain that when this star  formation rate is
employed the  fraction of  WDMS binaries climbs  to 31\% of  the total
synthetic  population.  However,  even  though the  number of  objects
redder than usual increases -- so  we obtain a better fit to the color
distribution  (see the  bottom panel  of Fig.~\ref{f:mod8sfr})  -- the
white   dwarf   luminosity   function   (see   the   left   panel   of
Fig.~\ref{f:mod8sfr}) is clearly shifted towards fainter magnitudes --
the   respective  reduced   $\chi^2$  values   are:  $\chi^2_{\nu,{\rm
    CD}}=1.34$ and $\chi^2_{\nu,{\rm LF}}=2.41$.  This excess of faint
white dwarfs prevents us to adopt  models of the Galactic bulge with a
significant fraction  of young objects.  Actually, we  conducted a set
of additional simulations to explore the maximum possible contribution
of  this  putative young  population  and  found  that the  sample  of
\cite{Calamida2014} restricts its contribution  to a modest 5\% of the
standard old population. This is the  same to say that the mass of the
young population  should be less  than 33\% of  the total mass  of the
Galactic bulge. In summary,  our results indicate that this population
is not only unable to explain the presence of anomalous red objects in
the  sample of  bulge  white dwarfs,  but  also is  at  odds with  the
observed white  dwarf luminosity function, unless  its contribution is
almost negligible.

\begin{figure}
\includegraphics[width=1.13\columnwidth]{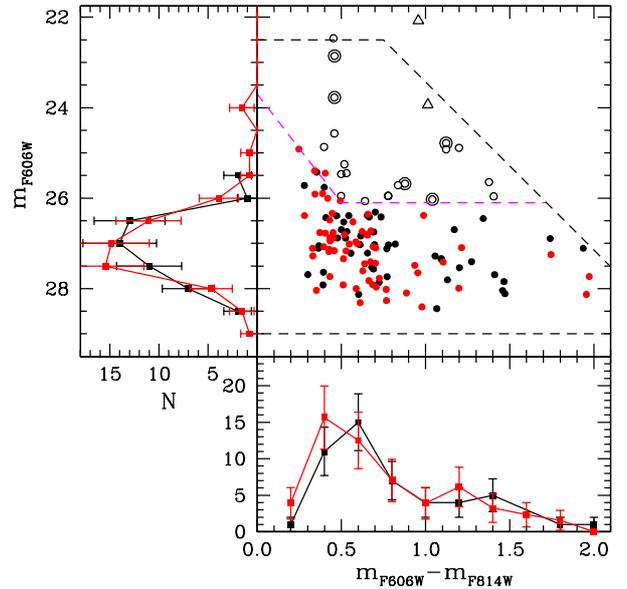}
   \caption{Same as Fig.~\ref{f:fig2} but for the population of single
     and  binary white dwarfs  of the  Galactic bulge.   Our simulated
     sample (red symbols and red lines) is that of Model~8, but adding
     differential  reddening  and  $10\%$  of  blended  objects.   The
     exclusion  region of  variable objects  is shown  using  a dashed
     magenta line.}
\label{f:bestfit}
\end{figure}

\subsection{Differential reddening and blending}
\label{other}

In previous sections  we have established that the  population of WDMS
is the main  contributor to the observed population  of red objects of
the Galactic bulge sample.  However,  even when a model that maximizes
the presence  of binaries is employed, some  discrepancies between the
simulated and observed sample persist. In the following we analyze two
possibilities  that  could help  in  alleviating these  discrepancies,
namely differential reddening  and blending.  Nevertheless, we advance
that we  foresee that the inclusion  of these two effects  will have a
minor impact on the results.

First, we include in Model~8  differential reddening. It is well known
that dense molecular clouds  towards the Galactic center exist.  These
molecular      clouds      produce      a      spatially      variable
extinction. Disentangling  the contribution of  differential reddening
for  each  individual  object  in  the  observed  sample  is  a  quite
unfordable   task,  which  is   beyond  the   scope  of   the  present
study. However,  we can estimate the effect  of differential reddening
using our Monte Carlo simulated samples.  To do this we follow closely
the   work   of   \cite{Calamida2014}.   Specifically,   to   simulate
differential  reddening  in  the  Sagittarius window,  we  assume  the
presence  of  random   clumps  of  full  width  at   half  maximum  of
$\approx5\,$arcsec (in total 500 clumps  fit in the ACS images) with a
reddening     increase      of     $E(F606W-F814W)=0.228$     ($A_{\rm
  F606W}=1.4291\,$mag)  on   the  top   of  a  uniform   reddening  of
$E(F606W-F814W)=0.5766$.

In addition to the possible  effects of differential reddening we also
consider  that  a  small  fraction  of objects  belong  to  unresolved
pairs. This is a reasonable  assumption, since the observed fields are
crowded.  In  particular, white  dwarf artificial tests  indicate that
the  percentage of  blended  stars could  be  as high  as $\sim  10\%$
\citep{Calamida2014}, which is the value we adopt here.

The results  for our best model (Model~8)  when differential reddening
and   blending  effects   are  taken   into  account   are   shown  in
Fig.~\ref{f:bestfit}.  To perform the analysis of the simulated sample
we  exclude from  the white  dwarf luminosity  function and  the color
distribution  those   objects  that  presumably   are  variable.   The
boundaries of the corresponding  region in the color-magnitude diagram
are  represented  in  Fig.~\ref{f:bestfit}  employing  magenta  dashed
lines.  Clearly, the agreement  between the observed (black lines) and
simulated  results  (red  lines)  is  now noticeably  good.   That  is
quantitative corroborate by reduced $\chi^2$ values close to 1 for the
luminosity  function  and  the color  distribution,  $\chi^2_{\nu,{\rm
    LF}}=1.05$ and $\chi^2_{\nu,{\rm  CD}}=1.16$, respectively.  It is
worth saying  that, even though differential  reddening and unresolved
pairs are  not by their own able  to explain the redder  region of the
observed color-magnitude diagram of the sample of \cite{Calamida2014},
their inclusion in our models clearly improves the fit.
 
\section{Conclusions}
\label{sec:conc}

Using  an state-of-the-art  Monte Carlo  simulator which  includes the
population of single  and binary white dwarfs and  the most up-to-date
evolutionary sequences,  we have  studied the observational  sample of
the Galactic bulge of \cite{Calamida2014}.  Our reference model with a
canonical    ratio   of   white    dwarfs   with    hydrogen-rich   to
hydrogen-deficient atmospheres reproduces fairly well the single white
dwarf population of  the galactic bulge, but is  not able to reproduce
other features of the color-magnitude diagram, and in particular could
not reproduce  the observed excess  of red objects.   Consequently, we
sought for  other alternatives. We  first analyzed the  possibility of
these  objects being  members of  a population  of  single helium-core
white dwarfs.  However, we found that the observed excess of red stars
in  the   color-magnitude  diagram   cannot  be  attributed   to  this
population.  We  then studied the  possibility of these  objects being
members  of  binaries.  We  found  that  our  Monte Carlo  simulations
including the binary white dwarf population reproduce better the gross
features of the color-magnitude  diagram, there are still some objects
that  cannot be  reproduced  by  the population  of  white dwarf  plus
main-sequence binaries,  because their colors  are bluer and  they are
fainter  than their  observed  counterparts, nor  by  a population  of
double degenerate  binaries, because its contribution  is smaller than
$2\%$, too small to account  for the observed number of objects. Thus,
the only  explanation we are  left with is  that these bright  and red
objects are  cataclysmic variables  or dwarf novae.   This explanation
agrees  with the  hypothesis put  forward by  \cite{Calamida2014}, who
found  that  some  of  these objects  present  variability.   However,
follow-up  observations   will  be   needed  to  fully   confirm  this
scenario. Once,  the putative  population of cataclysmic  variables is
eliminated from  the color-magnitude diagram  we obtain a nice  fit to
both   the  white  dwarf   luminosity  function   and  to   the  color
distribution,  when  a  fraction  $\la  10\%$ of  blended  objects  is
included in  the models.  The  best fit is obtained  when $\alpha_{\rm
  CE}=1$,   $\alpha_{\rm    int}=0.1$,   $n(q)=q^{-1}$,   $\alpha_{\rm
  IMF}=-2.55$   and  $f_{\rm   BIN}\simeq  0.80$   are   adopted,  and
differential reddening and blending  are included.  Although we cannot
rule out  the contribution of  young and thick-disk  metallicity bulge
stars,  we placed an  upper limit  to their  contribution to  the mass
budget of  the Galactic bulge,  and we found  that it must  be smaller
than $33\%$ of the total bulge mass.

\section*{Acknowledgements}
This work was partially supported by MINECO grant AYA2011--23102.

\bibliographystyle{mnras} \bibliography{bulb}

\bsp
\label{lastpage}
\end{document}